\documentclass[conference,letterpaper]{IEEEtran}

\usepackage[letterpaper, left=0.632in, right=0.632in, bottom=1in, top=0.75in]{geometry}

\usepackage{cite}

\usepackage{graphicx}
\usepackage{psfrag}
\usepackage{subfig}
\usepackage{overpic}
\usepackage{enumitem,kantlipsum}

\usepackage[cmex10]{amsmath}
\usepackage{amssymb}
\usepackage{color}
\usepackage{amsthm}

\def\tr{\mathrm{tr}}

\newtheorem{lemma}{Lemma}

\def\CN{\mathcal{N}_{\mathbb{C}}} %Complex Gaussian distribution
\newcommand{\vect}[1]{\mathbf{#1}}
\def\tr{\mathrm{tr}}

\def\Psiv{\vect{Q}}

\def\tr{\mathrm{tr}}

\def\Htran{\mbox{\tiny $\mathrm{H}$}}
\def\Ttran{\mbox{\tiny $\mathrm{T}$}}
\def\CN{\mathcal{N}_{\mathbb{C}}} %Complex Gaussian distribution
\def\taupu{\tau_{p}} %Uplink pilots
\def\bphiu{\boldsymbol{\phi}} %Notation of uplink pilot sequences
\usepackage[dvipsnames,svgnames]{xcolor}
\usepackage[textwidth=30mm]{todonotes}

\begin{document}

\IEEEoverridecommandlockouts

\IEEEoverridecommandlockouts

\title{What is the Benefit of Code-domain NOMA in Massive MIMO?\vspace{-.2cm}}
\author{
\IEEEauthorblockN{Mai T. P. Le\IEEEauthorrefmark{1}\IEEEauthorrefmark{4}, Luca Sanguinetti\IEEEauthorrefmark{2}, Emil Bj{\"o}rnson\IEEEauthorrefmark{3}, Maria-Gabriella Di Benedetto\IEEEauthorrefmark{1} \bigskip
%\thanks{This research has been supported by ELLIIT, the EU FP7 under ICT-619086 (MAMMOET) and the ERC Starting Grant 305123 MORE.}
\vspace{-.4cm}}
\IEEEauthorblockA{\IEEEauthorrefmark{1}\small{Department of Information Engineering, Electronics and Telecommunications, Sapienza University of Rome, Rome, Italy }}
\IEEEauthorblockA{\IEEEauthorrefmark{2}\small{Dipartimento di Ingegneria dell'Informazione, University of Pisa, Pisa, Italy}}
\IEEEauthorblockA{\IEEEauthorrefmark{3}\small{Department of Electrical Engineering (ISY), Link\"{o}ping University, Link\"{o}ping, Sweden}}
\IEEEauthorblockA{\IEEEauthorrefmark{4}\small{Faculty of Electronic and Telecommunication Engineering, The University of Danang - University of Science and Technology, Vietnam \vspace{-.4cm}}}}

\maketitle

\begin{abstract}

In overloaded Massive MIMO systems, wherein the number $K$ of user equipments (UEs) exceeds the number of base station antennas $M$, it has recently been shown that non-orthogonal multiple access (NOMA) can increase performance. This paper aims at identifying cases of the classical operating regime $K<M$, where code-domain NOMA can also improve the spectral efficiency of Massive MIMO. Particular attention is given to use cases in which poor favorable propagation conditions are experienced. Numerical results show that Massive MIMO with planar antenna arrays can benefit from NOMA in practical scenarios where the UEs are spatially close to each other.
\end{abstract}

%\begin{IEEEkeywords}
%Massive MIMO, uniform linear array, planar rectangular array, spatial correlation matrices, code-domain NOMA, spectral efficiency, channel estimation.
%\end{IEEEkeywords}

% Introduction
%!TEX root = PIMRC19_v4.tex

\section{Introduction}
Massive MIMO (mMIMO) \cite{massivemimobook} and Non-Orthogonal Multiple Access (NOMA) \cite{Dai2018Survey,LeTWC2018} are two physical layer technologies that have received large attention in recent years. While mMIMO has already made it into the 5G standard \cite{Parkvall2017a}, the NOMA functionality remains to be standardized. Since mMIMO will likely be a mainstream feature in 5G networks, it is important to determine if and how NOMA can improve its performance. Recent results in this direction can be found in \cite{Senel2019,Kuda2019NOMAaided, Liu2019MPANOMA}.

{Conventional multiple access schemes assign orthogonal resources to each user equipment (UE). This provides restricted/dedicated resources per UE but eliminates inter-UE interference. As well-known, this approach is inefficient if the interference can be controlled in some other domain \cite{Dai2018Survey,LeIET2018}; the power and code domains are typically used for interference suppression in the NOMA literature, while the spatial domain is used in mMIMO.} Although most of the previous work focuses on only one of these three domains, some recent works have considered the combination of power-domain NOMA in mMIMO systems \cite{Senel2019,Kuda2019NOMAaided,Zhang2017}. The gains are, however, generally limited since power-domain NOMA requires UEs with non-orthogonal channels to be efficient, while a core feature of mMIMO is to make the UE channels nearly orthogonal \cite{Senel2019}.

This paper addresses the potential combination of code-domain NOMA and mMIMO, that has received limited attention so far. In \cite{Ma2017NOMA}, the spectral efficiency (SE) of a code-domain NOMA technique, called interleave division multiple-access, was analyzed with an iterative data-aided channel estimation receiver.
In \cite{Liu2019MPANOMA}, it was shown that the SE of mMIMO can be improved by code-domain NOMA in the overloaded setting, where the number $K$ of active UEs in each cell is higher than the number $M$ of BS antennas, i.e., $M<K$. In contrast, this paper considers the traditional mMIMO regime wherein $M>K$.

As a matter of fact, the SE of a classical mMIMO system grows without bound as $M \to \infty$ when the spatial correlation properties of the interfering UEs' channels are sufficiently different \cite{BjornsonHS17,Sanguinetti_MassiveMIMO20}. Nevertheless, the SE that is achieved at any finite $M$ can potentially be improved. In particular, there are important use cases where the UEs are located close to each other, such as in public hubs like stadiums, offices in high-rise buildings, train stations, and in public outdoor events, wherein UEs' spatial channel correlation properties can be very similar and, thus, a very large number of antennas is needed to deliver acceptable performance when relying solely on the spatial processing provided by classical mMIMO. We will show that code-domain NOMA with judiciously designed spreading codes can provide the necessary additional degrees of freedom to manage interference between UEs with similar channels.

% In this specific setting, it is necessary to have additional degrees of freedom to separate UEs at the receiver. Code-domain NOMA, hence, may increase the spectral efficiency by assigning different codes to the UEs that are not resolvable in the solely mMIMO system.  

The paper is organized as follows. Section~\ref{Sec:SystemModel} introduces the multicell system model of mMIMO-NOMA. An achievable uplink SE and its optimal receive combining are derived in Section~\ref{Sec:SE}. Numerical results are provided in Section~\ref{Sec:NumericalAnalysis}, while conclusions are drawn in Section~\ref{Sec:conclusion}.

% System and Signal Model
%!TEX root = PIMRC19_v4.tex

\section{System Model}\label{Sec:SystemModel}
We consider an mMIMO network composed of $L$ cells. The BS in each cell is equipped with $M$ antennas and simultaneously serves $K$ single-antenna UEs. We assume that the BSs and UEs operate according to a time-division-duplex (TDD)
protocol with a data transmission phase and a pilot phase for channel estimation. We consider the standard block fading TDD protocol  \cite[Ch.~2]{massivemimobook} in which each coherence block consists of $\tau_c$ channel uses, whereof $\tau_p$ are used for uplink pilots, $\tau_u$ for uplink data, and $\tau_d$ for downlink data, with $\tau_c = \tau_p + \tau_u + \tau_d$. Only the uplink is considered in this work. We denote by $\vect{h}_{lk}^{j} \in \mathbb{C}^{M}$ the channel between UE~$k$ in cell~$l$ and BS~$j$. In each coherence block, an independent correlated Rayleigh fading realization is drawn:

\begin{equation} \label{eq:correlated-Rayleigh-model}
\vect{h}_{lk}^{j} \sim \CN \left( \vect{0}_{M}, \vect{R}_{lk}^{j}  \right)
\end{equation}
where $\vect{R}_{lk}^{j} \in \mathbb{C}^{M \times M}$ is the spatial correlation matrix. It describes the macroscopic propagation conditions and is known at the BS. The Gaussian distribution models the small-scale fading variations.
The normalized trace $\beta_{lk}^{j} =  \tr ( \vect{R}_{lk}^{j})/M$
%\begin{equation} \label{eq:beta-definition-lkj}
%\beta_{lk}^{j} = \frac{1}{M} \tr \left( \vect{R}_{lk}^{j} \right)
%\end{equation}
is the average channel gain from BS~$j$ to UE~$k$ in cell~$l$. 

\subsection{Channel estimation}

The uplink pilot sequence of {UE}~$k$ in cell~$j$ is denoted by $\bphiu_{jk} \in \mathbb{C}^{\taupu}$ and satisfies $\| \bphiu_{jk} \|^2  = \taupu$. The elements of $\bphiu_{jk}$ are scaled by the pilot power $\sqrt{p_{jk}}$ and transmitted over $\taupu$ channel uses, giving the received signal $\vect{Y}_j^{p} \in \mathbb{C}^{M \times \taupu}$ at {BS}~$j$:\begin{align} \label{eq:uplink-pilot-model}
\vect{Y}_j^{p} = \underbrace{ \sum_{i=1}^{K} \sqrt{p_{ji}} \vect{h}_{ji}^{j} \bphiu_{ji}^{\Ttran}  }_{\textrm{Desired pilots}} + \underbrace{\sum_{l=1,l \neq j}^{L} \sum_{i=1}^{K} \sqrt{p_{li}}  \vect{h}_{li}^{j} \bphiu_{li}^{\Ttran}  }_{\textrm{Inter-cell pilots}} + \underbrace{ \vphantom{\sum_{l=1,l \neq j}^{L} } \vect{N}_{j}^{p}}_{\textrm{Noise}}
\end{align}
where $\vect{N}_{j}^{p} \in \mathbb{C}^{M \times \taupu}$ is noise with i.i.d.\ elements distributed as $\CN(0,\sigma^{2})$. Note that we are not assuming mutually orthogonal pilot sequences, but arbitrary spreading sequences. Hence, the minimum mean-squared error (MMSE) estimator of $\vect{h}_{jk}^{j}$ takes a more complicated form than in prior work \cite[Ch.~3]{massivemimobook}.
%the MMSE estimator at BS~$j$ of the channel $\vect{h}_{li}^{j}$ of an arbitrary UE $i$ in cell $l$ can be computed as follows. 
\begin{lemma} \label{theorem:MMSE-estimate_h_jli}
The MMSE estimate of $\vect{h}_{li}^{j}$ is
\begin{equation} \label{eq:MMSEestimator_h_jli}
\widehat{\vect{h}}_{li}^{j}  = \sqrt{p_{li}}\left({\bphiu}_{li}^{\Htran} \otimes {\vect{R}}_{li}^j\right) \big(\Psiv_{li}^{j}\big)^{-1}  {\rm{vec}}\left({\bf Y}_{j}^{p}\right)
\end{equation}
where $\vect{Y}_j^{P}$ is given in \eqref{eq:uplink-pilot-model} and 
\begin{equation} \label{eq:Psiv-definition}
\Psiv_{li}^{j} =  \sum_{l^\prime=1}^{L}\sum_{i^\prime=1}^{K}  p_{l^\prime i^\prime}\left({\bphiu}_{l^\prime i^\prime}{\bphiu}_{l^\prime i^\prime}^{\Htran} \right)\otimes {\vect{R}}_{l^\prime i^\prime}^j+ \sigma^{2}{\bf I}_{M\tau_p}.
\end{equation}
The estimation error $\tilde{\vect{h}}_{li}^{j} = \vect{h}_{li}^{j} - \widehat{\vect{h}}_{li}^{j}$ is independent of $\widehat{\vect{h}}_{li}^j$ and has correlation matrix $
\vect{C}_{li}^{j} = \mathbb{E} \{ \tilde{\vect{h}}_{li}^{j} ( \tilde{\vect{h}}_{li}^{j} )^{\Htran} \}  = \vect{R}_{li}^{j} - \vect{\Phi}_{li}^j
$ with 
\begin{equation}\label{eq:Phi-definition1}
\vect{\Phi}_{li}^j ={p_{li}}\left({\bphiu}_{li}^{\Htran} \otimes {\vect{R}}_{li}^j\right)
{(\Psiv_{li}^{j})}^{-1}  \left({\bphiu}_{li} \otimes {\vect{R}}_{li}^j\right).
\end{equation} 
\end{lemma}
\begin{IEEEproof}
The proof follows from standard results and is omitted for space limitation.
\end{IEEEproof}
Notice that the MMSE estimate in \eqref{eq:MMSEestimator_h_jli} holds for any choice of pilot sequences $\{\bphiu_{li}\}$, which can be arbitrarily taken from an orthogonal, random or sparse set. In classical mMIMO, orthogonal pilot sequences are usually employed, leading to the simplified expressions found in \cite[Ch.~3]{massivemimobook}.

\subsection{Uplink data transmission}

While classical mMIMO only uses spreading sequences for uplink pilot transmission, the proposed mMIMO-NOMA system utilizes $N$-length spreading sequences also for the uplink data transmission.
We denote by $\vect{u}_{jk} \in \mathbb{C}^{N}$ the spreading sequence assigned to {UE}~$k$ in cell~$j$ and assume that $\| \vect{u}_{jk} \|^2  = N$. {As for pilot transmission, the spreading sequences $\{\vect{u}_{jk}\}$ are also taken from an arbitrary set.} The received signal ${\bf Y}_{j}\in \mathbb{C}^{M\times N}$ at BS $j$ is given by
\begin{align} \label{eq:uplink-signal-model}
\vect{Y}_j = \underbrace{ \sum_{i=1}^{K} s_{ji}\vect{h}_{ji}^{j}\vect{u}_{ji}^{\Ttran}}_{\textrm{Intra-cell signals}} + \underbrace{\sum_{l=1,l \neq j}^{L} \sum_{i=1}^{K}  s_{li}\vect{h}_{li}^{j}\vect{u}_{li}^{\Ttran}}_{\textrm{Inter-cell interference}} + \underbrace{\vphantom{\sum_{i=1,i\ne k}^{K} } \vect{N}_{j}}_{\textrm{Noise}}
\end{align}
where $s_{li} \sim \CN({0}, p_{li})$ is the data signal from UE~$i$ in cell~$l$ with $p_{{li}}$ being the transmit power and $\vect{N}_{j} \in \mathbb{C}^{M \times N}$ is thermal noise with i.i.d.\ elements distributed as $\CN(0, \sigma^{2})$. 
\begin{figure*}
\begin{align}\notag
\gamma_{jk}^{\rm {ul}} 
& =  \frac{ p_{jk}|  \vect{v}_{jk}^{\Htran} \widehat{\vect{g}}_{jk}^j |^2  }{{\mathbb{E}}\left\{ 
\!\sum\limits_{l=1,l\ne j}^L\sum\limits_{i=1}^K p_{li}| \vect{v}_{jk}^{\Htran} {\vect{g}}_{li}^j |^2 + \sum\limits_{i=1,i\ne k}^K p_{ji}| \vect{v}_{jk}^{\Htran} {\vect{g}}_{ji}^j |^2
+p_{jk}| \vect{v}_{jk}^{\Htran} \tilde{\vect{g}}_{jk}^j |^2+  \sigma^{2}\vect{v}_{jk}^{\Htran} \vect{v}_{jk} 
\Big| \{\widehat{\bf{h}}_{li}^{j} : \forall l,i\} \right\}} \notag\\&
= \frac{ p_{jk}|  \vect{v}_{jk}^{\Htran} \widehat{\vect{g}}_{jk}^j |^2  }{ 
 \vect{v}_{jk}^{\Htran}  \left(   \sum\limits_{l=1,l\ne j}^L\sum\limits_{i=1}^K p_{li}\widehat{\vect{g}}_{li}^j {(\widehat{\vect{g}}_{li}^j)}^{\Htran}+\sum\limits_{i=1,i\ne k}^K p_{ji}\widehat{\vect{g}}_{jk}^j {(\widehat{\vect{g}}_{jk}^j)}^{\Htran} +   {\vect{Z}}_j \right) \vect{v}_{jk}  
}   \label{eq:uplink-instant-SINR}\tag{10}
\end{align}
\hrule
\begin{align} 
&\left[ \vect{R}_{li}^j \right]_{m_1,m_2} =\beta_{li}^j \iint{ \underbrace{e^{\mathsf{j} \pi (m_1-m_2) \sin(\theta)}}_{\textrm{Vertical correlation}} \underbrace{e^{\mathsf{j} \pi (m_1-m_2) \cos(\theta)\sin(\varphi)}}_{\textrm{Horizontal correlation}} f(\varphi,\theta) d\varphi d\theta }
\label{eq:3DChannelModel}\tag{15}\vspace{-0.45cm}
\end{align}
\hrule
\end{figure*}

% Spectral Efficiency Analysis and Receive Combining
%!TEX root = PIMRC19_v4.tex

\section{Spectral Efficiency}\label{Sec:SE}
%We analyze the achievable SE of the maMIMO-NOMA in the uplink. 
To detect the data signal $s_{jk}$ from $\vect{Y}_j$ in \eqref{eq:uplink-signal-model}, BS~$j$ selects the combining vector $\vect{v}_{jk} \in \mathbb{C}^{MN}$, which is multiplied with the vectorized version of $\vect{Y}_j$ to obtain
\begin{align} \notag
\!\!\vect{v}_{jk}^{\Htran} {\rm{vec}}\left({\bf Y}_{j}\right) 
%&=  \sum_{l=1}^{L} \sum_{i=1}^{K}  s_{li}\vect{v}_{jk}^{\Htran}{\rm{vec}}\left(\vect{h}_{li}^{j}\vect{u}_{li}^{\Htran}\right) + \vect{v}_{jk}^{\Htran}{\rm{vec}}\left(\vect{N}_j\right)
%\\\notag
&=  s_{jk} \vect{v}_{jk}^{\Htran}\vect{g}_{jk}^{j} + \underbrace{ \sum_{i=1,i\ne k}^{K} s_{ji}\vect{v}_{jk}^{\Htran}\vect{g}_{ji}^{j}}_{\textrm{Intra-cell interference}} \\&+ \underbrace{\sum_{l=1,l \neq j}^{L} \sum_{i=1}^{K}s_{li} \vect{v}_{jk}^{\Htran}\vect{g}_{li}^{j}}_{\textrm{Inter-cell interference}} + \underbrace{\vphantom{\sum_{i=1,i\ne k}^{K} } \vect{v}_{jk}^{\Htran}{\rm{vec}}\left({\bf N}_{j}\right)}_{\textrm{Noise}}\label{eq:vector-channel-UL-processed}
\end{align}
where $\vect{g}_{li}^{j}  = {\rm{vec}}\big(\vect{h}_{li}^{j} \vect{u}_{li}^{\Htran}\big) \in \mathbb{C}^{MN}$ or, equivalently,
\begin{align}\label{effective_channel_vec}
\vect{g}_{li}^{j} =\vect{u}_{li} \otimes \vect{h}_{li}^{j}=\left(\vect{u}_{li} \otimes {\bf I}_{M}\right)\vect{h}_{li}^{j}
\end{align}
is the \emph{effective channel vector} with correlation matrix $\mathbb{E} \{ \vect{g}_{li}^{j} (\vect{g}_{li}^{j} )^{\Htran} 
\} = \left(\vect{u}_{li} \otimes {\bf I}_{M}\right)\vect{R}_{li}^{j} \left(\vect{u}_{li}^{\Htran} \otimes {\bf I}_{M}\right)$.
The MMSE estimate of $\vect{g}_{li}^{j}$ is obtained as $\widehat{\vect{g}}_{li}^{j} = \vect{u}_{li} \otimes \widehat{\vect{h}}_{li}^{j}=\left(\vect{u}_{li} \otimes {\bf I}_{M}\right)\widehat{\vect{h}}_{li}^{j}$.

%and has correlation matrix 
%\begin{align}\notag
%\mathbb{E} \{ \widehat{\vect{g}}_{li}^{j} ( \widehat{\vect{g}}_{li}^{j} )^{\Htran} \}  &= \left(\vect{u}_{li} \otimes {\bf I}_{M}\right){\vect{\Phi}_{li}^{j}} \left(\vect{u}_{li}^{\Htran} \otimes {\bf I}_{M}\right)\\&=\left(\vect{u}_{li}\vect{u}_{li}^{\Htran}\right) \otimes{\vect{\Phi}}_{li}^{j}.
%\end{align}
The ergodic capacity under imperfect CSI is generally unknown, but there exist well-established lower bounds that can be used to rigorously analyze performance \cite{massivemimobook}.

\begin{lemma}\label{theorem:uplink-SE}
If the MMSE estimator is used, an uplink SE of UE $k$ in cell $j$ is
\begin{equation} \label{eq:uplink-rate-expression-general}
\begin{split}
\mathsf{SE}_{jk}^{\rm {ul}} = \frac{1}{N}\frac{\tau_u}{\tau_c} \mathbb{E} \left\{ \log_2  \left( 1 + \gamma_{jk}^{\rm {ul}}  \right) \right\} \quad \textrm{[bit/s/Hz] }
\end{split}
\end{equation}
where the effective instantaneous signal-to-interference-and-noise ratio (SINR) $\gamma_{jk}^{\rm {ul}}$ is given in \eqref{eq:uplink-instant-SINR} (see next page)
%, where ${\mathbb{E}}\{\cdot|\{\widehat{\bf{h}}_{li}^{j}: \forall l,i\}\}$ denotes the conditional expectation given the {MMSE} channel estimates $\{\widehat{\bf{h}}_{li}^{j}: \forall l,i\}$ at BS~$j$, and
and
\setcounter{equation}{10}
\begin{align}
{\vect{Z}}_j = \sum_{l=1}^{L} \sum_{i=1}^{K} p_{li}\left(\vect{u}_{li}\vect{u}_{li}^{\Htran}\right) \otimes\vect{C}_{li}^{j} +\sigma^{2} {\bf I}_{MN}.\end{align}
\end{lemma}
\begin{IEEEproof}
The proof follows the same steps as the proof of \cite[Th. 4.1]{massivemimobook} and is therefore omitted.
\end{IEEEproof}
The pre-log factor $\frac{1}{N}\frac{\tau_u}{\tau_c}$ accounts for the fraction of samples used for uplink data. It is smaller than $\frac{\tau_u}{\tau_c}$, which would be the case with classical mMIMO (i.e., in the absence of spreading sequences for data transmission), but if the sequences are properly associated to the UEs, the SINR can be substantially higher.

The SE expression in \eqref{eq:uplink-rate-expression-general} holds for any combining vector and choice of spreading sequences for data transmission. A possible choice for $\vect{v}_{j k}$ is to use maximum ratio (MR) combining with $\vect{v}_{j k} = \widehat{\vect{g}}_{jk}^j$. However, \eqref{eq:uplink-instant-SINR} has the form of a generalized Rayleigh quotient. Hence, the vector that maximizes it can be obtained as follows.

\begin{lemma} The {SINR} in \eqref{eq:uplink-instant-SINR} is maximized by
\begin{align} \label{eq:MMSE-combining}
\vect{v}_{j k} =p_{jk}\Bigg(  \sum\limits_{l=1}^L \sum\limits_{i=1}^Kp_{li}\widehat{\vect{g}}_{li}^j {(\widehat{\vect{g}}_{li}^j)}^{\Htran} + \vect{Z}_j  \Bigg)^{\!-1}    \widehat{\vect{g}}_{jk}^j
\end{align}
which leads to 
\begin{align} \label{eq:MMSE-max-SINR}
\gamma_{jk}^{\rm {ul}}  &= p_{jk}{(\widehat{\vect{g}}_{jk}^j)}^{\Htran}\!\!\left(  \!\sum\limits_{(l,i)\neq (j,k)}\!\!\!\!p_{li}\widehat{\vect{g}}_{li}^j {(\widehat{\vect{g}}_{li}^j)}^{\Htran} + \vect{Z}_j  \right)^{\!-1}   \!\! \!\!\widehat{\vect{g}}_{jk}^j.
\end{align}
\end{lemma}
\begin{IEEEproof}
This result follows from \cite[Lemma B.10]{massivemimobook}.
\end{IEEEproof}

{The combining vector $\vect{v}_{j k}$ in \eqref{eq:MMSE-combining} is a function of the effective MMSE estimates $\{\widehat{\vect{g}}_{jk}^j\}$, rather than $\{\widehat{\vect{h}}_{jk}^j\}$ as would be the case in classical mMIMO.}

It can be shown that the SINR-maximizing combiner in \eqref{eq:MMSE-combining} minimizes the mean-squared error ${\rm{MSE}}_k = \mathbb{E} \{ | s_{jk} - \vect{v}_{jk}^{\Htran} {\rm{vec}}\left({\bf Y}_{j}\right)   |^2  \big| \{ \hat{\vect{h}}_{li}^j \}  \}$, which represents the conditional MSE between the data signal $s_{jk}$ and the received signal $\vect{v}_{jk}^{\Htran} {\rm{vec}}\left({\bf Y}_{j}\right)$ after
receive combining; see \cite[Sec.~4.1]{massivemimobook} for details. We will therefore call it \emph{NOMA multicell-MMSE (M-MMSE) combining}.
The ``multicell'' notion refers to the fact that it is computed by utilizing both the intra- and inter-cell channel estimates that can be computed locally at BS $j$, using the existing pilot signaling. No cooperation between the cells is needed. Compared to heuristic solutions, such as NOMA-MR combining with $\vect{v}_{jk} = \widehat{\vect{g}}_{jk}^j$, it has higher computational complexity since it requires first the computation of the $MN\times MN$ matrix inverse in \eqref{eq:MMSE-combining} and then a matrix-vector multiplication. In comparison, the computational complexity of MR is $MN$ complex multiplications for channel use; see \cite[Sec.~4.1.2]{massivemimobook} for further details and alternative heuristic solutions.

% Numerical Analysis
%!TEX root = PIMRC19_v4.tex
\begin{table*}[t]
\renewcommand{\arraystretch}{1.}
\centering
\caption{Network parameters}
\label{table:system_parameters_running_example}
\begin{tabular}{|c|c|}
\hline \bfseries $\!\!\!\!\!$ Parameter $\!\!\!\!\!$ & \bfseries Value\\
\hline\hline

%  Network layout & $\!$Random deployment (with wrap-around)$\!$ \\
      Cell size &  $250$\,m $ \times\,  250$\,m \\
%    Distance of cluster with its BS $d_{\textrm{BSclus}}$ & $100$\,m  \\
% %    Radius of cluster  & $[7.8, 15.6, 31.25]$\,m  \\
%   UE dropping  &      \begin{tabular}{@{}c@{}} $K$ UEs are randomly distributed within a cluster \\with 35m minimum distance with its BS  \end{tabular}\\
%% Bandwidth & $B = 20$\,MHz  \\
%BS antennas & $M =64$\\
UL noise power and UL transmit power & $\sigma^2 = -94$\,dBm, $p_{jk}= 20$\,dBm \\
%
%UL transmit power  & $\rho_{\rm{ul}}= 20$\,dBm\\

Samples per coherence block & $\tau_c = 200$ \\

%Pilot reuse factor & $f=1$ \\

Distance between UE $i$ in cell $l$ and BS~$j$ & $d_{li}^{\,j}$ \\

\begin{tabular}{@{}c@{}} Large-scale fading coefficient for \\the channel between UE $i$ in cell $l$ and BS~$j$\end{tabular}
& $\beta_{li}^{j} =  -148.1 - 37.6 \, \log_{10} \left( \frac{d_{li}^{j}}{1\,\textrm{km}} \right) + F_{li}^{j}$\,dB\\
 
Shadow fading between UE $i$ in cell $l$ and BS~$j$ & $F_{li}^{j} \sim \mathcal{N}(0,10)$ \\

%Angular standard deviation (ASD) & $\sigma_{\varphi}^2= 2^\circ$ \\

% Pathloss exponent & $\alpha = 3.76$ \\

%$\!$Shadow fading (standard deviation)$\!$  & $\sigma_{\mathrm{sf}}= 10$ \\

\hline
\end{tabular}\vspace{-0.6cm}
\end{table*}
\section{Numerical analysis}\label{Sec:NumericalAnalysis}
To quantify the benefits of code-domain NOMA in mMIMO, we numerically evaluate the SE for the network setup in Table \ref{table:system_parameters_running_example} using Lemma~\ref{theorem:uplink-SE}. Two different antenna geometries and channel models are considered, as follows:

\begin{enumerate}
\item The 2D one-ring channel model for a uniform linear array with half-wavelength spacing and average path loss $\beta_{li}^j$ \cite[Sec. 2.6]{massivemimobook}. For an angle-of-arrival (AoA) $\varphi_{li}^j$, the scatterers are uniformly distributed in $[\varphi_{li}^j -\Delta, \varphi_{li}^j + \Delta]$ with $\Delta$ being the angular spread. This makes the $(m_1,m_2)$th element of $\vect{R}_{li}^j$ be:
\begin{align}\label{eq:2DChannelModel}
\left[ \vect{R}_{li}^j \right]_{m_1,m_2} =\frac{\beta_{li}^j}{2\Delta} \int_{-\Delta}^{\Delta}{ e^{\mathsf{j} \pi(m_1-m_2) \sin(\varphi_{li}^j + {\varphi}) }}d{\varphi}.
\end{align}
\item The 3D one-ring channel model for a planar array with the antennas uniformly spaced with half-wavelength horizontal and vertical spacing \cite[Sec. 7.3]{massivemimobook}. In this case, the $(m_1,m_2)$th element of $\vect{R}_{li}^j$ is given by \eqref{eq:3DChannelModel} (see top of this page), where $f(\varphi,\theta) $ is the joint PDF of the azimuth $\varphi$ and elevation $\theta$ angles. 
%$j(m_1)-j(m_2)$ and $i(m_1)-i(m_2)$ stand for the vertical and horizontal distance between antennas $\textrm{m}_1$ and $\textrm{m}_2$, respectively. Due to space constraints, readers are referred to \cite{massivemimobook} for further details. 
We consider a planar array consisting of $\sqrt M$ horizontal rows with $\sqrt M$ antennas each. Following \cite[Fig. 7.14, Sec. 7.3.2]{massivemimobook}, the 3D model is implemented by assuming that the BS height is $25$ m, the UE height is $1.5$ m, and a uniform angular distribution is used. 

\end{enumerate}
The analysis is carried out with MR and M-MMSE combining. When mMIMO-NOMA is employed, $\vect{v}_{jk}$ is thus given by $\vect{v}_{jk} = \widehat{\vect{g}}_{jk}^j$ and \eqref{eq:MMSE-combining}, respectively. We assume that a set of $N$ orthogonal codes of length $N$ is used and randomly associated to the UEs. When mMIMO is considered, $\vect{v}_{jk} = \widehat{\vect{h}}_{jk}^j$ with MR and is given by \cite[Eq. (32)]{BjornsonHS17} with M-MMSE. {Due to space limitation, we only consider orthogonal spreading codes for both pilot and data transmission. The case of either random or sparse spreading sequences is left for the extended version.}

\subsection{A single-cell two-user scenario}
We begin by considering the simple setup with $L=1$, $M=64$, and $K=2$. %Particularly, we assume that the UEs are distributed randomly within a cluster with radius $r = 1$ m. This implies that the two UEs have very similar AoAs $\varphi_{11}^1$ and $\varphi_{12}^1$ to the BS. This is a challenging setup with possibly poor favorable propagation, and it will clearly showcase the main use cases wherein NOMA can bring some benefits.
{To investigate the SE behavior of UEs with respect to their locations, we fix the nominal angle of one UE at $30^{\circ}$ and let the nominal angle of the second one vary from $-180^{\circ}$ to $180^{\circ}$.}

\begin{figure}[t!]
\begin{center}
\subfloat[2D channel model.]{\label{SE_funcOfAoI_1cell2UEs_2D}
\begin{overpic}[unit=1mm,width=1\columnwidth]{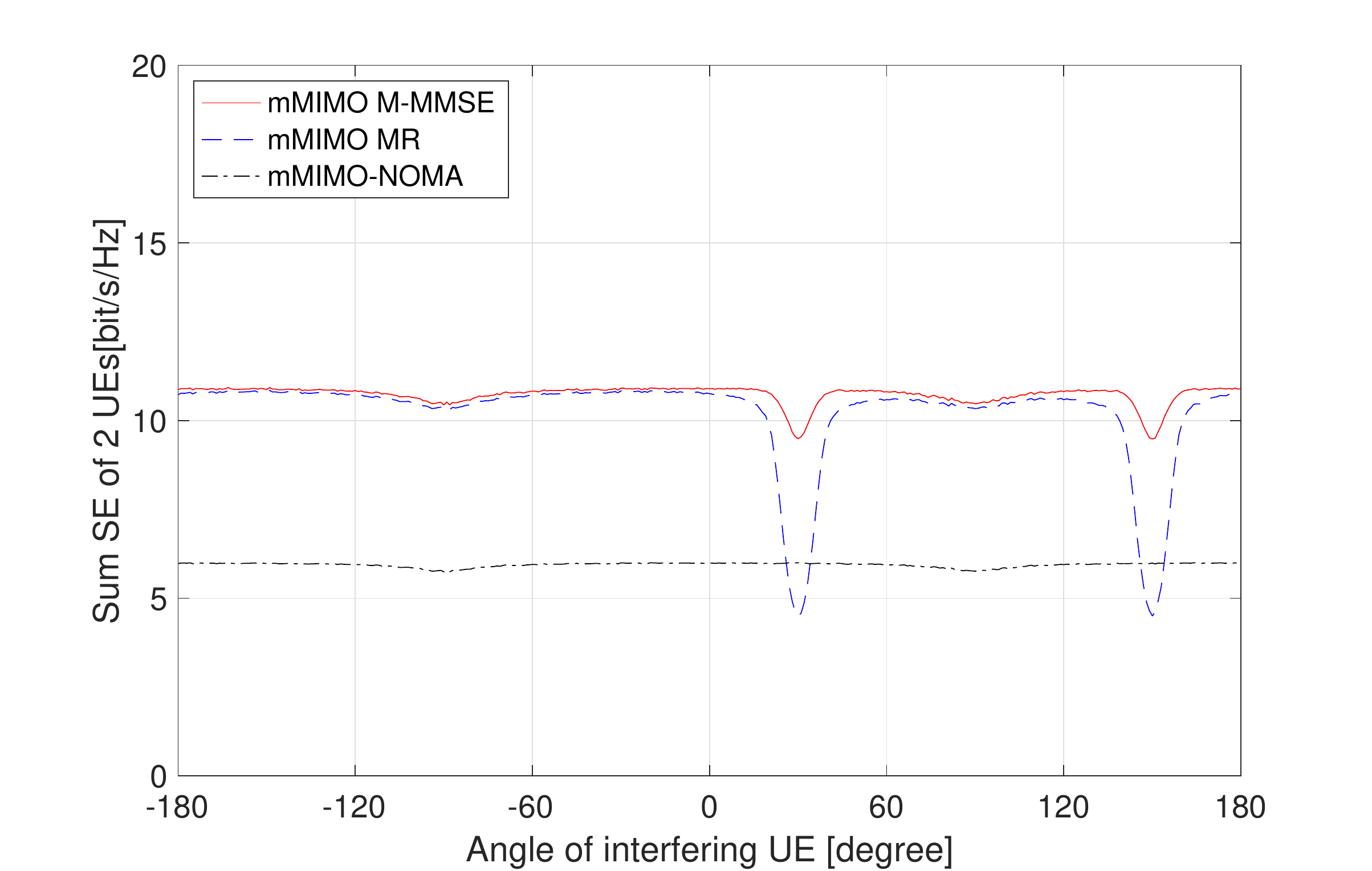}
\end{overpic}} \\
\vspace{-0.35cm}
\subfloat[3D channel model.]{\label{SE_funcOfAoI_1cell2UEs_3D}
\begin{overpic}[unit=1mm,width=1\columnwidth]{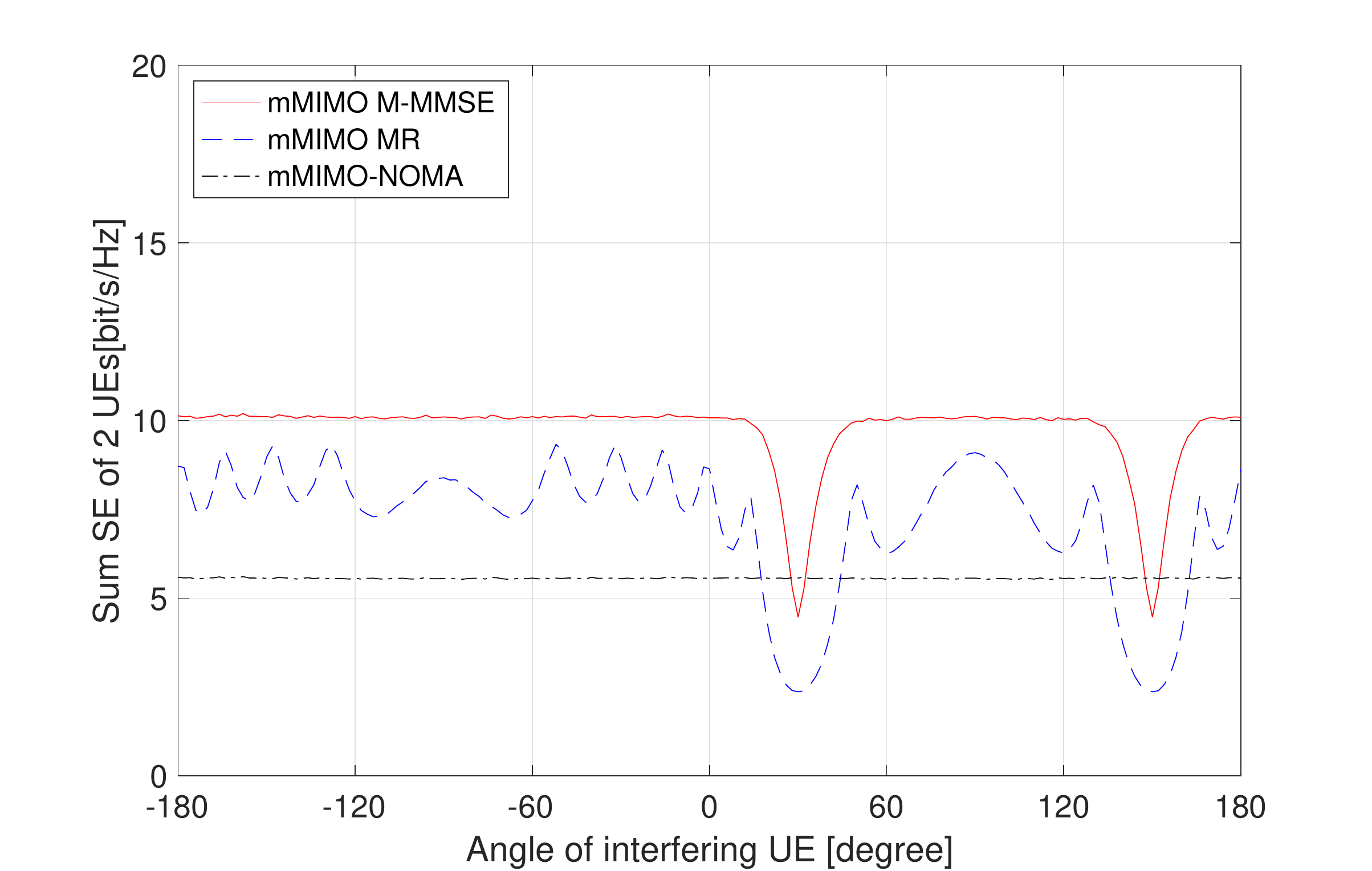}
\end{overpic}} 
\end{center} 
\vspace{-0.35cm}
\caption{{Average sum SE for a single-cell two-user setup with $\Delta=2^{\circ}$ and $M = 64$ with mMIMO and mMIMO-NOMA for $N=2$, assuming the nominal angle of the desired UE is fixed at $30^{\circ}$, while the angle of the interfering UE ranges from $-180^{\circ}$ to $180^{\circ}$. The 2D and 3D channel models are considered.}}\label{SE_funcOfAoI_1cell2UEs}
\vspace{-0.25cm}
\end{figure}

Fig.~\ref{SE_funcOfAoI_1cell2UEs} shows the average sum SE of the two UEs with classical mMIMO and mMIMO-NOMA. In the latter scheme, M-MMSE and MR perform exactly the same since $N=K=2$ and thus no interference is present---this is why only one curve is reported with mMIMO-NOMA. Both channel models are considered with a relatively small ASD $\Delta = 2^{\circ}$ and with $M=64$. {Fig.~\ref{SE_funcOfAoI_1cell2UEs} shows that classical mMIMO gives higher SE than NOMA in both 2D and 3D models for most of the interfering angles. Different results are obtained for the case in which the interfering UE has a very similar AoA to the BS. This is a challenging setup characterized by poor favorable propagation, wherein NOMA can bring some benefits. For the 2D model, Fig.~\ref{SE_funcOfAoI_1cell2UEs}(a) shows that  M-MMSE largely outperforms NOMA even in this poor favorable propagation condition.} This is because M-MMSE is a sufficiently powerful scheme to reject the interference even when the UEs are very close in space. However, we notice that this is achieved at the cost of a higher computational complexity \cite{massivemimobook}. Specifically, it scales as $(NM)^3$ with M-MMSE, rather than as $NM$ with NOMA. Fig.~\ref{SE_funcOfAoI_1cell2UEs} shows also that NOMA can provide some gain compared to MR, without any increase of the complexity. For the 3D model, Fig.~\ref{SE_funcOfAoI_1cell2UEs}(b) reveals that NOMA provides the highest SE irrespective of the combining scheme used with mMIMO when UEs are close in space. This is because the array has a smaller spatial resolution, which reduces the interference rejection capabilities of mMIMO in the spatial domain.

\begin{figure}[t!]
\centering\vspace{0cm}
\includegraphics[unit=1mm,width=1\columnwidth]{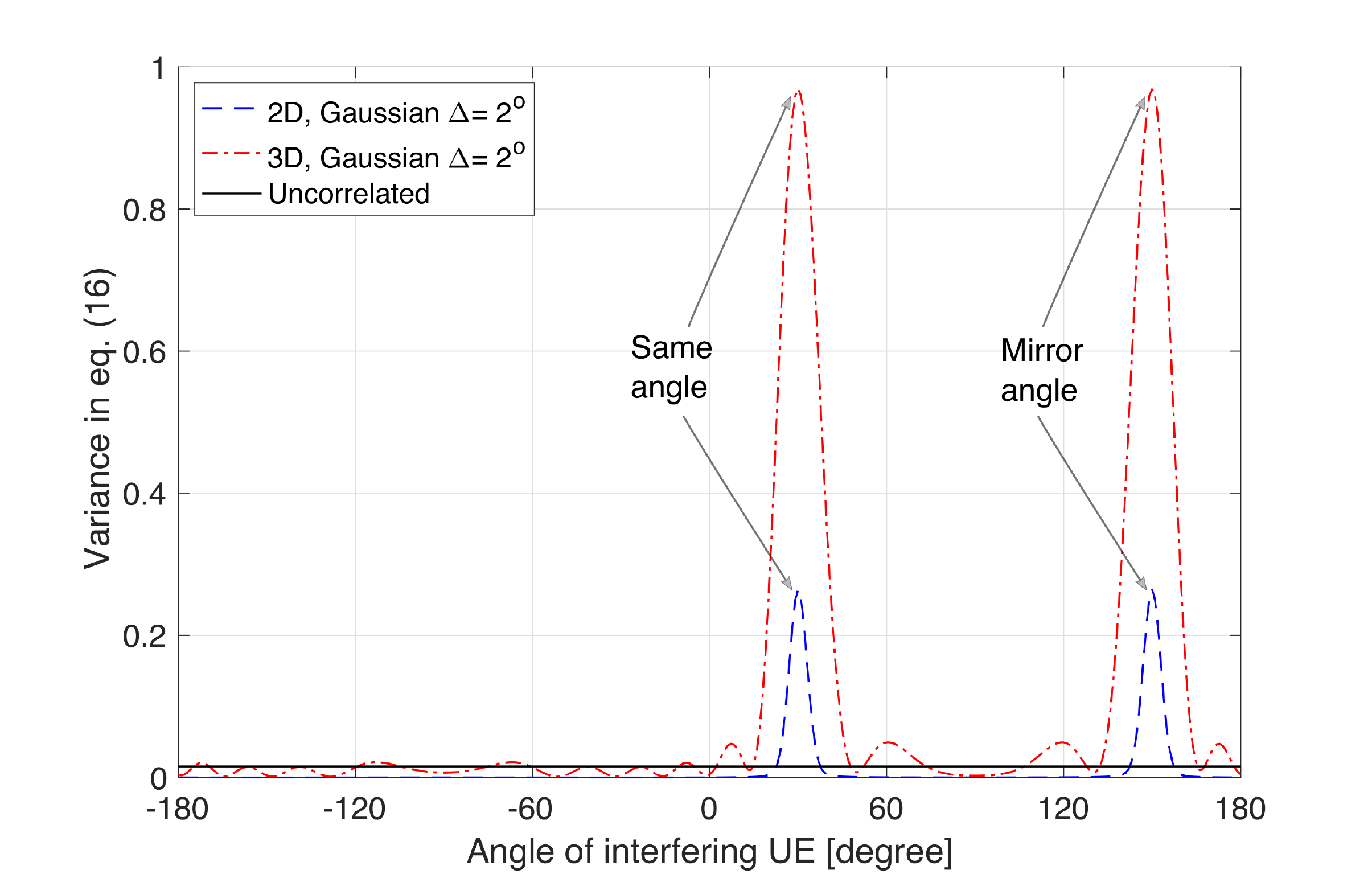}
\caption{{Behaviour of the variance defined in \eqref{eq:variance-favorable-propagation} for the same setup of Fig. 1. Uncorrelated fading is also reported for comparison with the 2D and 3D channel models.}}\vspace{-0.35cm}
\label{fig:figVar_funcOfAoI}
\end{figure}

% Fig.~\ref{SE_funcOfM_1cell2UEs_3D_ASD5} considers the same setup of Fig. 1(b) but with a higher angular spread $\Delta = 5^{\circ}$. In this case, M-MMSE provides the highest SE, followed by NOMA and then MR. This indicates that having a lower spatial correlation (i.e., higher angular spread) makes the spatial correlation matrices of the UEs sufficiently different to be separated with M-MMSE. This happens even if the UEs are characterized by very similar AoAs. However, this is not sufficient with MR, which is always a worse choice than NOMA.

%By using orthogonal codes, the sum SE for M-MMSE and MR combining techniques in maMIMO-NOMA are equal to each other, and higher than both cases of maMIMO, despite the number of BS antennas. It is interesting to observe a similar behavior for `3D channel' with planar array, even with higher ASD, such as $\sigma_{\varphi} = 2^{\circ}$. This is because the spatial resolution of the planar array are now shared for azimuth and elevation angular directions, i.e.\ there are only 8 antennas are dedicated for azimuth resolution instead of 64 as the ULA case, leading to the decrease in SE performance for 3D channels.   
To better understand these results, Fig.~\ref{fig:figVar_funcOfAoI} plots the variance
\setcounter{equation}{15}
\begin{equation} \label{eq:variance-favorable-propagation}
\begin{split}
\mathbb{V}  \left \{ 
\frac{(\vect{h}_{11}^1)^{\Htran}\vect{h}_{12}^{1}}{\sqrt{\mathbb{E} \{ \| \vect{h}_{11}^1 \|^2 \} \mathbb{E} \{ \| \vect{h}_{12}^{1}  \|^2 \} } }
\right \}
= \frac{\tr \left( \vect{R}_{11}^1 \vect{R}_{12}^1 \right) }
 { \tr (\vect{R}_{11}^1) \tr ( \vect{R}_{12}^1 ) }
\end{split}
\end{equation}
of the two UEs for 2D and 3D models in the setup of Fig.~\ref{SE_funcOfAoI_1cell2UEs}, which measures the so-called \textit{favorable propagation} condition \cite[Eq. (2.19)]{massivemimobook}. This is a measure of how much interference the two UEs cause to each other. A smaller variance corresponds to lower interference.
As seen, it is a function of the spatial correlation matrices of the UEs. Ideally, the variance should be zero. {Fig.~\ref{SE_funcOfAoI_1cell2UEs} shows that \eqref{eq:variance-favorable-propagation} achieves its maximum values at $30^{\circ}$ and $150^{\circ}$ for both channel models. These correspond to the SE drops in Fig.~\ref{SE_funcOfAoI_1cell2UEs}. With the 2D model, the variance peaks are relatively small ($\approx 0.25$), leading to comparatively good favorable propagation conditions. This justifies why classical mMIMO performs well in the setup of Fig.~\ref{SE_funcOfAoI_1cell2UEs}(a). On the other hand, the variance is substantially larger ($\approx 0.95$) with the 3D model. This is because the array has only a horizontal spatial resolution given by 8 antennas and a vertical spatial resolution given by 8 antennas, thus it cannot separate the users in any of these domains.} Hence, the two UEs cause much interference to each other, and thus the SE of mMIMO deteriorates, especially with MR. This issue can be solved by using NOMA, as shown previously in Fig.~\ref{SE_funcOfAoI_1cell2UEs}(b).

\subsection{A multi-cell scenario with clusterized UEs}

%\begin{figure}[t!]
%\centering
%\includegraphics[unit=1mm,width=1\columnwidth]{4cell_KUEclusters}
%\caption{Illustration of 4-cell setup of the running simulation. In each $0.25 \times 0.25$km cell is located a BS with either a vertical ULA $M=64$ antennas, or a planar rectangular array consisting of $M=8\times8$ antennas in the center. $K$ UEs per cell focus into clusters. The size and position of the UE cluster can be varied randomly such that the system requirements in Table~\ref{table:system_parameters_running_example} are satisfied. }
%\label{fig:figSetup2}
%\end{figure}

We now extend the SE analysis to a more general scenario with $L=4$ cells. Each BS is equipped with $M = 64$ antennas and serve $K$ UEs that are gathered together in a cluster of radius $r$. Each cluster is randomly dropped within its own cell. As for the previous scenario, this setup is quite challenging for conventional mMIMO since the spatial resolution is often insufficient to separate the users, which may result in poor favorable propagation conditions between the UEs. We stress that there are important use cases where this scenario may occur, for example, in stadiums, train stations, public events, and so forth. With NOMA, each cluster is divided into $K/N$ (which is assumed to be an integer) subclusters, each with $N$ UEs. We assume that $N$ orthogonal codes are randomly assigned to the $N$ UEs that belong to a given subcluster.

%With NOMA, each cluster is divided into $K/N$ (which is assumed to be an integer-valued) subclusters, each with $N$ UEs. We assume that $N$ orthogonal codes are randomly assigned to the $N$ UEs that belong to a given subcluster. In doing so, mMIMO-NOMA allows to improve the SE performance, though it introduces an $N$ times lower pre-log factor in the SE expression given in \ref{eq:uplink-rate-expression-general}. The impact of $K$ and $r$ on the SE performance of mMIMO and mMIMO-NOMA are investigated for further insights.

\begin{figure}[t!]
\begin{center}
\subfloat[2D model.]{\label{SE_funcOfK_random_2D}
\begin{overpic}[unit=1mm,width=1\columnwidth]{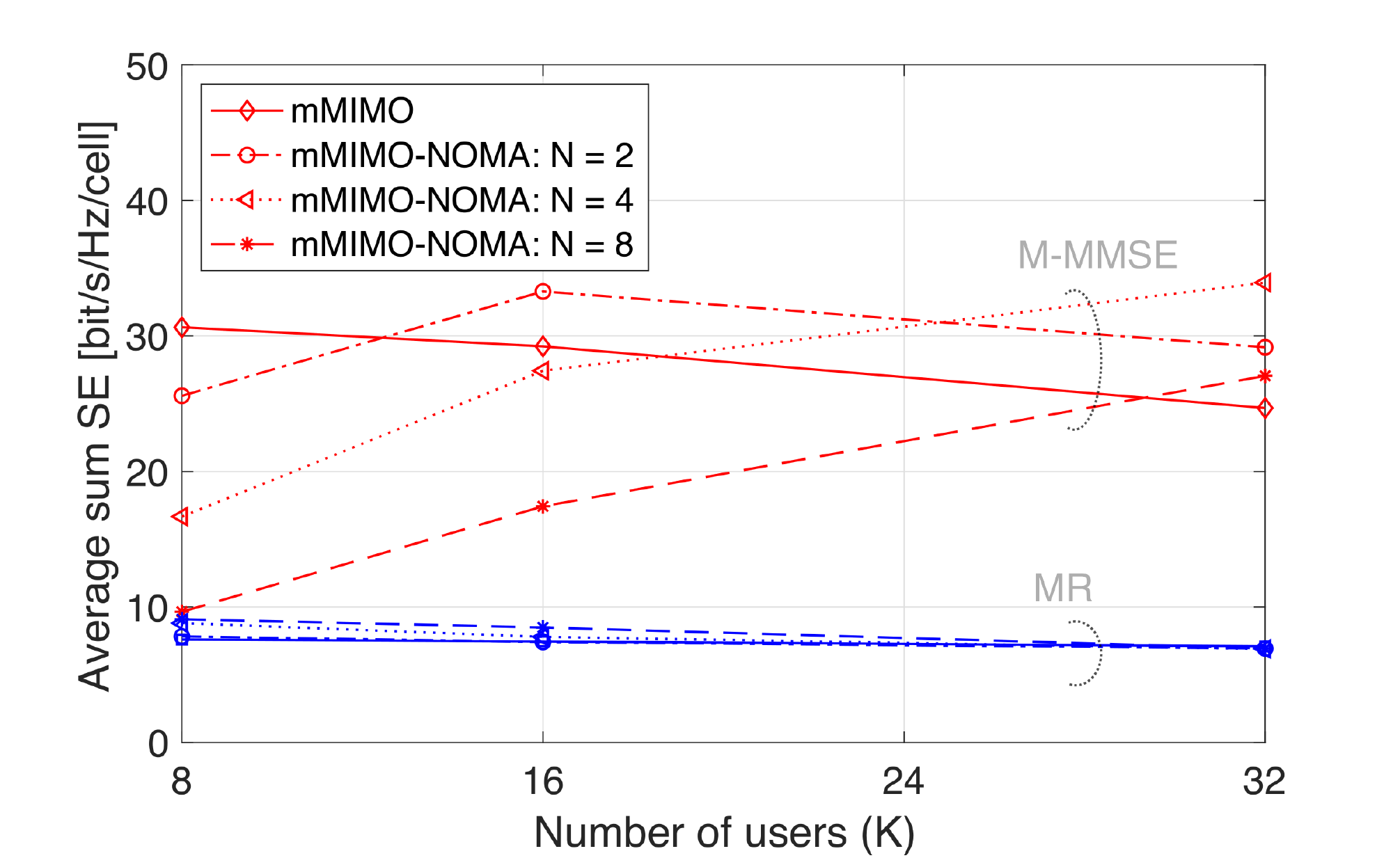}
\end{overpic}} \\
\vspace{-0.35cm}
\subfloat[3D model.]{\label{SE_funcOfK_random_3D}
\begin{overpic}[unit=1mm,width=1\columnwidth]{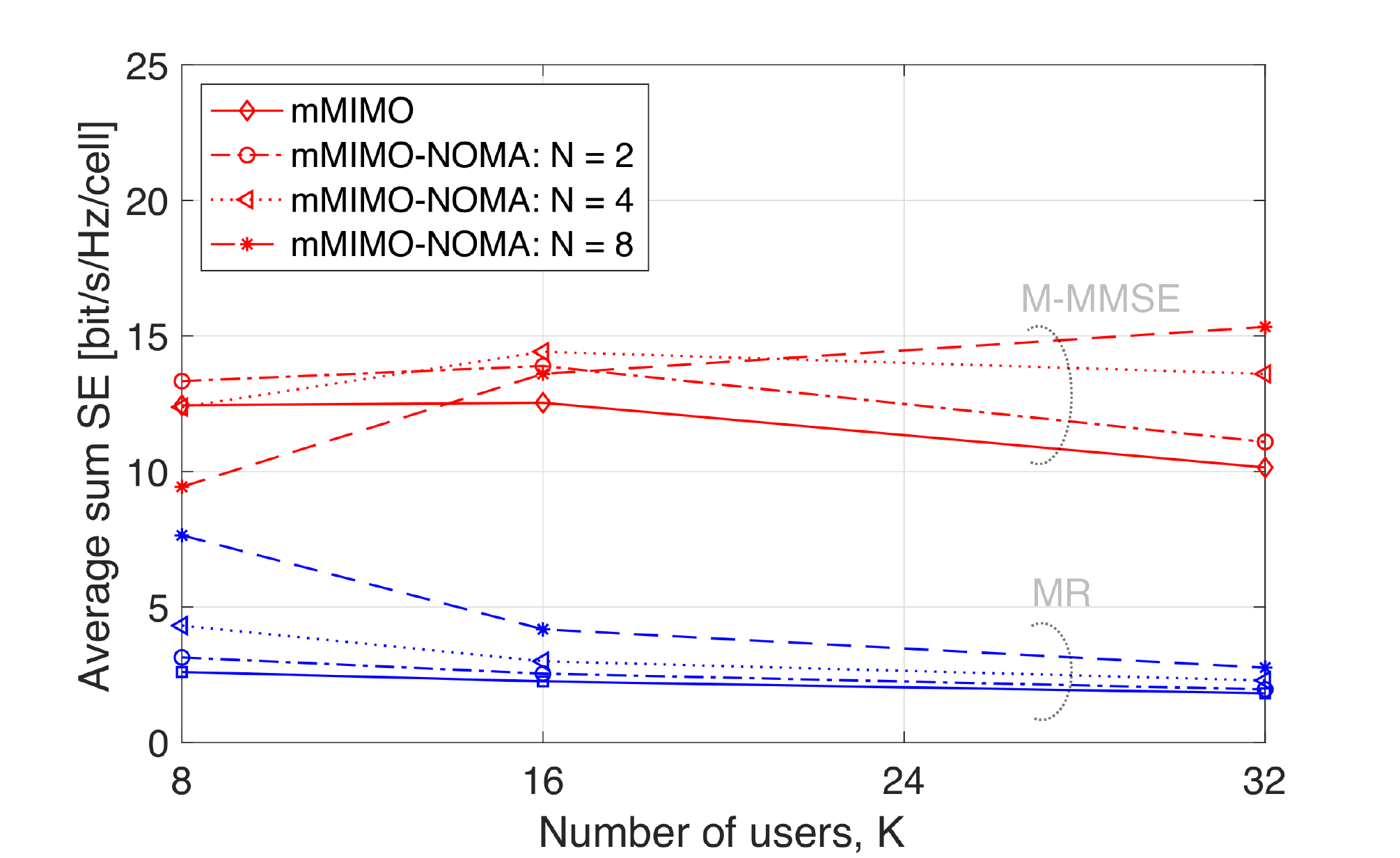}
\end{overpic}} 
\end{center}
\vspace{-0.35cm}
\caption{Average sum SE for the multi-cell $K$-users setup with $\Delta=2^{\circ}$ vs. $K$ with mMIMO and mMIMO-NOMA for $N \in \{2,4,8\}$. The 2D and 3D models are both considered.}\label{SE_funcOfK_random}\vspace{-0.45cm}
\end{figure}

Figure ~\ref{SE_funcOfK_random} shows the average sum SE per cell as a function of $K$ for mMIMO and mMIMO-NOMA with $N \in \{2,4,8\}$. M-MMSE and MR are considered with both 2D and 3D channel models, and with the angular spread $\Delta = 2^{\circ}$. The cluster radius is $r=10$ m and each cluster is randomly positioned at a distance larger than $25$~m from its serving BS. %It is found that the effective number of codes in mMIMO-NOMA corresponds to the number of users that is needed to be resolved in the angular spatial domain. %Noting that the number of BS antennas is $M=64$ for the ULA, and $M = M_H \times M_V = 8 \times 8$ for the planar array, where $M_H$ and $M_V$ are the number of antennas in horizontal and vertical dimensions. Thus the number of antennas to resolve UEs in azimuth angles for `3D channel' strongly decrease compared to the `2D channel' case, leading to the fact that more codes  are necessary to assign to UEs for achieving higher SE performance of planar array. 
From Fig.~\ref{SE_funcOfK_random}, we see that the SE of M-MMSE with mMIMO-NOMA %with  and $N\ge 4$ increases with $K$ irrespective of the channel model. This happens even though a relative small angular spread $\Delta = 2^{\circ}$ is considered. Figure ~\ref{SE_funcOfK_random}(b) shows that mMIMO-NOMA 
achieves better performance than with mMIMO, provided that a good combination of $K$ and $N$ is considered. For example, a $25\%$ gain for 2D model and $35\%$ gain for 3D model are achieved with $N=4$ and $K=32$. This is because the 3D model has a lower resolution in the horizontal/vertical angular domains, which penalizes mMIMO---as also observed in previous work (c.f. \cite[Sec. 7.4]{massivemimobook}). Comparing the results of Figs.~\ref{SE_funcOfK_random}(a) and \ref{SE_funcOfK_random}(b) we observe that the SE of MR is very marginally affected by $N$ in the 2D case, while it increases with $N$ in the 3D case. This is consistent with the basic setup of Sec.~IV.A, where the interference between the UEs in the 3D scenario can be reduced by using NOMA. In summary, the above results show that NOMA allows to improve the SE of mMIMO in the considered cases. This happens even if the use of spreading sequences for uplink data transmission introduces an $N$ times lower pre-log factor in the SE expression in (\ref{eq:uplink-rate-expression-general}). 
 
%% When there are less interference, the SE performance with MR combining thus improves respectively. 
%
%\begin{figure}[t!]
%\centering
%\includegraphics[unit=1mm,width=1\columnwidth]{SE_FuncOfRadCluster_random}
%\caption{Average sum SE of $K = 16$ UEs for the multi-cell setup with $\Delta=2^{\circ}$ as a function of $r$ with mMIMO and mMIMO-NOMA for $N = 2$. The 2D and 3D channels are considered.}\vspace{-0.35cm}
%\label{fig:figFuncRadCluster}
%\end{figure}
%
%We now investigate the impact of the cluster radius $r$. We assume $K=16$ and $N=2$. Fig.~\ref{fig:figFuncRadCluster} shows that mMIMO and mMIMO-NOMA have the same performance with MR for any $r$. In the 2D case, it is preferable to employ NOMA with M-MMSE when $r \le 14$ m. With M-MMSE and 3D model, mMIMO-NOMA is always better choice than mMIMO but the gain is moderately marginal. 

% Conclusions
%!TEX root = PIMRC19_v4.tex
\section{Conclusions}\label{Sec:conclusion}
This paper investigated the potential benefits of code-domain NOMA in Massive MIMO systems. Particularly, we showed that the SE can be improved by using NOMA when poor favorable propagation conditions are experienced by the UEs. This may happen when they are located close to each other and the antenna array does not provide sufficient resolution in the spatial domain. {For example, numerical results were used to show that planar rectangular arrays with $M=64$ antennas can benefit from NOMA when the UEs experience poor favorable propagation conditions.} This is the type of arrays that are currently being deployed in 4G and 5G networks. These results were obtained both for the heuristic MR combiner and the optimal M-MMSE combiner. This shows that NOMA may help even if schemes with high interference rejection capabilities are employed.

\bibliographystyle{IEEEtran}
\bibliography{IEEEabrv,refs,ref,ref_book}

% Generated by IEEEtran.bst, version: 1.14 (2015/08/26)
\begin{thebibliography}{10}
\providecommand{\url}[1]{#1}
\csname url@samestyle\endcsname
\providecommand{\newblock}{\relax}
\providecommand{\bibinfo}[2]{#2}
\providecommand{\BIBentrySTDinterwordspacing}{\spaceskip=0pt\relax}
\providecommand{\BIBentryALTinterwordstretchfactor}{4}
\providecommand{\BIBentryALTinterwordspacing}{\spaceskip=\fontdimen2\font plus
\BIBentryALTinterwordstretchfactor\fontdimen3\font minus
  \fontdimen4\font\relax}
\providecommand{\BIBforeignlanguage}[2]{{%
\expandafter\ifx\csname l@#1\endcsname\relax
\typeout{** WARNING: IEEEtran.bst: No hyphenation pattern has been}%
\typeout{** loaded for the language `#1'. Using the pattern for}%
\typeout{** the default language instead.}%
\else
\language=\csname l@#1\endcsname
\fi
#2}}
\providecommand{\BIBdecl}{\relax}
\BIBdecl

\bibitem{massivemimobook}
\BIBentryALTinterwordspacing
E.~Bj\"{o}rnson, J.~Hoydis, and L.~Sanguinetti, ``Massive {MIMO} networks:
  {Spectral}, energy, and hardware efficiency,'' \emph{Foundations and
  Trends{\textregistered} in Signal Processing}, vol.~11, no. 3-4, pp.
  154--655, 2017. [Online]. Available:
  \url{http://dx.doi.org/10.1561/2000000093}
\BIBentrySTDinterwordspacing

\bibitem{Dai2018Survey}
L.~{Dai}, B.~{Wang}, Z.~{Ding}, Z.~{Wang}, S.~{Chen}, and L.~{Hanzo}, ``A
  survey of non-orthogonal multiple access for {5G},'' \emph{IEEE
  Communications Surveys Tutorials}, vol.~20, no.~3, pp. 2294--2323,
  thirdquarter 2018.

\bibitem{LeTWC2018}
M.~T.~P. {Le}, G.~C. {Ferrante}, T.~Q.~S. {Quek}, and M.~{Di Benedetto},
  ``Fundamental limits of low-density spreading {NOMA} with fading,''
  \emph{IEEE Transactions on Wireless Communications}, vol.~17, no.~7, pp.
  4648--4659, July 2018.

\bibitem{Parkvall2017a}
S.~Parkvall, E.~Dahlman, A.~Furusk\"ar, and M.~Frenne, ``{NR}: The new {5G}
  radio access technology,'' \emph{IEEE Communications Standards Magazine},
  vol.~1, no.~4, pp. 24--30, Dec 2017.

\bibitem{Senel2019}
K.~{Senel}, H.~V. {Cheng}, E.~{Bj\"ornson}, and E.~G. {Larsson}, ``What role
  can {NOMA} play in massive {MIMO}?'' \emph{IEEE Journal of Selected Topics in
  Signal Processing}, vol.~13, no.~3, pp. 597--611, June 2019.

\bibitem{Kuda2019NOMAaided}
D.~{Kudathanthirige} and G.~A.~A. {Baduge}, ``{NOMA}-aided multicell downlink
  massive {MIMO},'' \emph{IEEE Journal of Selected Topics in Signal
  Processing}, vol.~13, no.~3, pp. 612--627, June 2019.

\bibitem{Liu2019MPANOMA}
L.~{Liu}, C.~{Yuen}, Y.~L. {Guan}, Y.~{Li}, and C.~{Huang}, ``Gaussian message
  passing for overloaded massive {MIMO-NOMA},'' \emph{IEEE Transactions on
  Wireless Communications}, vol.~18, no.~1, pp. 210--226, Jan 2019.

\bibitem{LeIET2018}
M.~T.~P. {Le}, G.~C. {Ferrante}, G.~{Caso}, L.~{De Nardis}, and M.~{Di
  Benedetto}, ``On information-theoretic limits of code-domain {NOMA for 5G},''
  \emph{IET Communications}, vol.~12, no.~15, pp. 1864--1871, 2018.

\bibitem{Zhang2017}
D.~{Zhang}, Z.~{Zhou}, C.~{Xu}, Y.~{Zhang}, J.~{Rodriguez}, and T.~{Sato},
  ``Capacity analysis of {NOMA} with {mmWave} massive {MIMO} systems,''
  \emph{IEEE Journal on Selected Areas in Communications}, vol.~35, no.~7, pp.
  1606--1618, July 2017.

\bibitem{Ma2017NOMA}
J.~{Ma}, C.~{Liang}, C.~{Xu}, and L.~{Ping}, ``On orthogonal and superimposed
  pilot schemes in massive {MIMO} {NOMA} systems,'' \emph{IEEE J. Sel. Areas
  Commun.}, vol.~35, no.~12, pp. 2696--2707, Dec 2017.

\bibitem{BjornsonHS17}
E.~Bj{\"o}rnson, J.~Hoydis, and L.~Sanguinetti, ``Massive {MIMO} has unlimited
  capacity,'' \emph{IEEE Trans. Wireless Commun.}, vol.~17, no.~1, pp.
  574--590, Jan. 2018.

\bibitem{Sanguinetti_MassiveMIMO20}
\BIBentryALTinterwordspacing
L.~Sanguinetti, E.~Bj\"ornson, and J.~Hoydis, ``Towards {Massive MIMO 2.0}:
  Understanding spatial correlation, interference suppression, and pilot
  contamination,'' \emph{CoRR}, vol. abs/1904.03406, 2019. [Online]. Available:
  \url{https://arxiv.org/abs/1904.03406}
\BIBentrySTDinterwordspacing

\end{thebibliography}

%\section{Appendix}
%Assume $M=1$. Then, we get
%\begin{align} 
%\vect{y} = s\vect{u}^{\Ttran} + \vect{n}^{\Ttran}\end{align}
%\begin{align} 
%\hat{s} = \vect{y} \vect{u} =  s\vect{u}^{\Ttran}\vect{u} + \vect{n}^{\Ttran}\vect{u}\end{align}
%
%\begin{align} 
%{\rm{SNR}} = \frac{{\rm{E}}\{|s\vect{u}^{\Ttran}\vect{u}|^2\}}{{\rm{E}}\{|\vect{n}^{\Ttran}\vect{u}|^2\}}= \frac{{\rm{E}}\{|s\vect{u}^{\Ttran}\vect{u}|^2\}}{\vect{u}^{\Ttran}{\rm{E}}\{\vect{n}\vect{n}^{\Ttran}\}\vect{u}}= \frac{\vect{u}^{\Ttran}\vect{u}^2{\rm{E}}\{|s|^2\}}{\vect{u}^{\Ttran}\sigma^2{\bf I}_N\vect{u}}= \frac{N^2p}{\sigma^2 N}=\frac{Np}{\sigma^2}\end{align}
%
%\begin{align} 
%{\rm{SE}} = \frac{1}{N}\log_2\left(1+ \frac{Np}{\sigma^2}\right)\end{align}
\end{document}